\documentclass[debug]{rmaa}

\usepackage{paralist}
\usepackage{amsmath}

\usepackage{psfrag,color}

\usepackage[latin1]{inputenc}

\title{Analytical solutions for the dynamical clock A$+$ indicator in a toy model of pure dynamical friction} 

\author{
  M. Pasquato,\altaffilmark{1,2}}

\altaffiltext{1}{INAF, Osservatorio Astronomico di Padova, vicolo dell'Osservatorio 5, I--35122 Padova, Italy.}
\altaffiltext{2}{INFN, Sezione di Padova, Via Marzolo 8, I--35131 Padova, Italy.}

\shortauthor{Pasquato}
\shorttitle{Analytical A$+$ indicator solutions}

\fulladdresses{
\item M. Pasquato, INAF, Osservatorio Astronomico di Padova, vicolo dell'Osservatorio 5, I--35122 Padova, Italy.
}

\listofauthors{M. Pasquato}
\indexauthor{Pasquato, M.}

\abstract{Blue straggler stars are more massive than the average star in globular clusters, as they originate from the merger of two stars. Consequently, they experience dynamical friction, progressively sinking to the cluster center. Recently, several indicators of the degree of dynamical relaxation of a globular cluster have been proposed, based on the observed radial distribution of blue straggler stars. The most successful is the Alessandrini indicator, or A$+$ for short, which is the integral of the cumulative distribution of the blue straggler stars minus that of a lighter reference population. A$+$ correlates with the dynamical age of a cluster both in realistic simulations and in observations. Here I calculate the temporal dependence of the A$+$ indicator analytically in a simplified model of the evolution of the blue straggler star distribution under dynamical friction only.}

\resumen{}

\addkeyword{Galaxy: globular clusters: general}
\addkeyword{Methods: analytical}
\addkeyword{Stars: blue stragglers}

\begin{document}
\maketitle

\section{Introduction}
\label{sec:intro}
Blue straggler stars are found in all globular clusters observed to date in the Milky way \citep[][]{2004ApJ...604L.109P}. They are heavier than the average star in their host clusters, as they originate from stellar mergers either through direct collision \citep[][]{1976ApL....17...87H} or close-binary mass transfer \citep[][]{1964MNRAS.128..147M, 2009Natur.457..288K} or both \citep[][]{2004MNRAS.349..129D, 2004ApJ...605L..29M}.
Since the first observations of a bimodality in the radial distribution of blue straggler stars when normalized to a reference population \citep[][]{1993AJ....106.2324F, 1997A&A...327.1004Z}, attempts at understanding its origin and evolution have been made based on simulations run with different software and various levels of realism \citep{2004ApJ...605L..29M, 2006MNRAS.373..361M, 2012Natur.492..393F, 2013MNRAS.429.1221H, 2015ApJ...799...44M, 2017MNRAS.471.2537H, 2019MNRAS.483.1523S}.

One of the main goals of these efforts was to reproduce the formation of a minimum in the normalized blue straggler star radial distribution, or \emph{zone of avoidance}, starting from initial conditions where the blue straggler progenitors (i.e. close binary stars if for the moment we exclude the direct collision channel) are not mass segregated, sharing the same distribution as the other stars in the cluster. A secondary goal was to predict the temporal evolution of the position of the minimum and other characteristics of their distribution.
In particular \citet{2012Natur.492..393F} interpreted the position of the minimum as the hand of a \emph{dynamical clock}, powered essentially by mass segregation of  blue straggler stars, revealing the state of dynamical relaxation of the host cluster. Further work based on direct N-body simulations \citep{2015ApJ...799...44M} and on Montecarlo simulations \citep{2017MNRAS.471.2537H} found the formation and motion of the blue straggler minimum somewhat hard to reproduce reliably due to a combination of statistical noise (as blue straggler stars are already low in number in real clusters, simulations necessarily contain an even lower number of particles that a real cluster, and the blue straggler number is even lower -by definition- in the minimum of their distribution) and possibly intrinsic dynamical noise. To solve the issue of statistical noise \citet{2019MNRAS.483.1523S} introduced a technique named \emph{artificial oversampling} where they simulate a large number of blue stragglers in a Montecarlo simulation while treating them as tracers, i.e. without letting them affect the dynamics of regular stars. Still, while  \citet{2019MNRAS.483.1523S} manage to reliably observe the formation of a minimum in the blue straggler radial distribution, their simulations encounter difficultes in reproducing the motion of the minimum in absolute physical units \citep[as expected by observations and found by][]{2012Natur.492..393F} even though the radial position of their minimum increases with respect to the core radius of the cluster, which is shrinking in physical units while the cluster moves toward core collapse.

In a previous paper \citet{2018ApJ...867..163P} showed that the physical ingredients underlying the formation and motion of the minimum are dynamical friction and diffusion respectively. While the two are connected as they ultimately arise from the same phenomenon, i.e. scatter with lighter background stellar particles, \citet{2018ApJ...867..163P} varied the diffusion coefficient and dynamical friction independently, showing that when diffusion is too strong a minimum does not reliably form, whereas if diffusion is too weak a clear-cut minimum forms but does not move outwards over time. 
This suggests that simulation schemes should be carefully assessed regarding to their ability to correctly model the dynamical friction and diffusion phenomena in order to reproduce the observed evolution of the blue straggler star distribution minimum with increasing dynamical age.

In this context \citet{2016ApJ...833..252A} introduced a new \emph{dynamical clock} indicator which did not require a measurement of the position of the minimum of the normalized blue straggler star distribution, as it is based on the cumulative radial distribution of blue straggler stars compared to the cumulative distribution of some other class of reference stars. The \citet{2016ApJ...833..252A} indicator (or A$+$ for short) was introduced in the context of direct N-body simulations, where it was shown that it increases with the dynamical age of simulated clusters, acting as a mass-segregation powered dynamical clock. Later, \citet{2016ApJ...833L..29L} measured (a slightly modified version of) the A$+$ indicator on a sample of $25$ Galactic globular clusters, showing that it correlates with the cluster dynamical age measured in terms of a cluster's current relaxation time.

The A$+$ indicator is defined as the difference between the integral of the cumulative distribution of the blue straggler stars, expressed as a function of the logarithm of the cluster-centric radius, and that of a reference distribution. In the following I will obtain some of its properties analytically under simplifying assumptions.

\section{Calculations}
\subsection{A toy model of dynamical friction}
I model blue straggler stars as a population of particles in circular orbits in a spherically symmetric fixed gravitational potential. The radius $r$ of each orbit evolves due to dynamical friction, as
\begin{equation}
\label{drift}
\dot{r} = - \frac{r}{\tau(r)} = - v(r),
\end{equation}
where $r$ is the distance from the center and $\tau(r)$ is a positive, monotonically increasing function of $r$, representing the scale friction time at radius $r$.

Equation \ref{drift} shows that orbit radii contract with an instantaneous velocity $v(r) > 0$ that depends only on $r$. It can be integrated, obtaining
\begin{equation}
\label{rimplicit}
\int_{r_0}^{r} \frac{\tau (x) dx}{x} = -t,
\end{equation}
where $r_0$ is the initial value of the radius at time $t = 0$ and $r$ is its current value at time $t$. In general $r_0 > r$ because the radii contract over time. If the function $\tau(x)$ is known, the integral can be calculated and $r$ can be obtained as a function of $r_0$ and $t$:
\begin{equation}
\label{rexplicit}
r(r_0, t) = I^{-1}(I(r_0) - t),
\end{equation}
where the primitive
\begin{equation}
\label{integral}
I(r) = \int \frac{\tau (x) dx}{x}
\end{equation}
is an invertible function because $\tau(x)/x$ always is positive. It is easy to see that $r(r_0, t)$ is a monotonically decreasing function of $t$ for every $t > 0$ and for every $r_0$, i.e. that orbit radii keep shrinking over time.
Similarly
\begin{equation}
\label{rexplicitrev}
r_0(r, t) = I^{-1}(I(r) + t),
\end{equation}
also holds.

I now denote with $N(r, t)$ the cumulative distribution of particles at a given time as a function of radius. This is by construction such that $N(0, t) = 0$ and $\lim_{r \to \infty} N(r, t) = 1$ for all $t$. If for any two particles at time $t=0$ the condition ${r_0}_1 < {r_0}_2$ held, then at any subsequent $t$, ${r_1(t)} < {r_2(t)}$ will also hold. Therefore
\begin{equation}
N(r, t) = N(r_0(r, t), 0)
\end{equation}
as the number of particles that had a radius less than a given $r_0$ at the beginning still have a radius less than $r(r_0, t)$ at time $t$. This can be rewritten as 
\begin{equation}
\label{cum}
N(r, t) = N(I^{-1}(I(r) + t), 0)
\end{equation}
which, given knowledge of the function $I$ is a general solution for $N(r,t)$. Thus $\tau(r)$ fully determines $N(r,t)$ given an initial $N(r,0)$.

\subsection{Recovering the A$+$ indicator}
In the following I will assume that the reference population of stars to which the blue stragglers are compared to build the A$+$ indicator initially shares the same distribution as the blue stragglers and does not evolve.

Under this assumption it is trivial to obtain the evolution of the (three-dimensional) A$+$ indicator from Eq.~\ref{cum}. I will write $s = \log r$, so that
\begin{equation}
\label{cumlog}
N(r, t) = N(I^{-1}(I(e^s) + t), 0)
\end{equation}
so the A$+$ indicator becomes
 \begin{equation}
\label{aless}
A^{+}(t) = \int_{-\infty}^{+\infty} N(I^{-1}(I(e^s) + t), 0) ds - \int_{-\infty}^{+\infty} N(e^s, 0) ds
\end{equation}

\subsection{Monotonicity}
Note that at time $t_2 > t_1$
 \begin{equation}
\label{alessdiff}
A^{+}(t_2) -  A^{+}(t_1) = \int_{-\infty}^{+\infty} \left[ N(I^{-1}(I(e^s) + t_2), 0) - N(I^{-1}(I(e^s) + t_1), 0) \right] ds
\end{equation}
and the integrand
 \begin{equation}
\label{alessdiff}
N(I^{-1}(I(e^s) + t_2), 0) - N(I^{-1}(I(e^s) + t_1), 0)
\end{equation}
is positive for every $s$, because $I^{-1}(I(e^s) + t_2) > I^{-1}(I(e^s) + t_1)$ as the two terms represent, per Eq.~\ref{rexplicitrev}, the initial radius of a particle that is at $r = e^s$ at $t_2$ and $t_1$ respectively: a particle that took more ($t_2 > t_1$) to fall to $r$ was further away at the beginning.
This implies that $A^{+}(t)$ is a monotonically increasing function of time, i.e. a working dynamical clock.

\subsection{A$+$ linear dependence in globular cluser cores}
While Eq.~\ref{rexplicitrev} can be solved numerically for any $\tau(r)$, some choices of $\tau(r)$ will lead to a simple analytical solution.
 For example, following Eq.~1 of \citet{2004ApJ...605L..29M} I take
 \begin{equation}
 \tau(r) \propto \frac{\sigma^3(r)}{\rho(r)}
 \end{equation}
 where $\sigma$ is the velocity dispersion of background stars at radius $r$ and $\rho$ is their number density.
 For a Plummer model this works out as
 \begin{equation}
  \label{tau}
 \tau(r) = \tau_0 \left( 1 + \frac{r^2}{a^2} \right)^{7/4}
\end{equation}
where $a$ is the model scale radius and $\tau_0$ the scale time for dynamical friction at the center. From Eq.~\ref{integral}, setting $\tau_0 = 1$ and $a=1$ I obtain
\begin{equation}
I(u) = \frac{1}{2} \left[ \log(u-1) - log(u + 1) \right] + \arctan(u) + \frac{2}{7} u^7 + \frac{2}{3} u^3
\end{equation}
where
\begin{equation}
u = \left( 1 + {r^2} \right)^{1/4} > 1
\end{equation}
which unfortunately cannot be inverted in terms of simple functions. 
However for small radii Eq.~\ref{tau} reduces to a constant, so Eq.~\ref{integral} becomes trivially
\begin{equation}
I(r) = \tau_0 \log(r/a)
\end{equation}
and
\begin{equation}
\label{expodep}
r = r_0 e^{-t/\tau_0}
\end{equation}
so
\begin{equation}
\label{cumconst}
N(r, t) = N(a e^{\log(r/a) + t/\tau_0}, 0) = N(r e^{t/\tau_0}, 0)
\end{equation}
As the central regions of a Plummer model have approximately constant density $\rho_0$, I can take
\begin{equation}
N_c(r, 0) = 4 \pi \rho_0 r^3
\end{equation}
with a radial cutoff at
\begin{equation}
{r_c}_0 = \left( 4 \pi \rho_0 \right)^{-1/3}
\end{equation}
after which $N_c(r, 0)$ becomes identically $1$.
At time $t$ the radius at which $N_c(r, t)$ becomes identically $1$ is
\begin{equation}
r_c = {r_c}_0 e^{-t/\tau_0}
\end{equation}
Therefore
\begin{equation}
\label{aless_const}
A^{+}(t) = \int_{-\infty}^{\log{r_c}} N_c(e^{s + t/\tau_0}, 0) ds + \int_{\log{r_c}}^{\log{{r_c}_0}} 1 ds - \int_{-\infty}^{\log{{r_c}_0}} N_c(e^s, 0) ds 
\end{equation}
which simplifies to
\begin{equation}
\label{aless_const_clean}
A^{+}(t) =  \log{\frac{{r_c}_0}{r_c}} = \frac{t}{\tau_0}.
\end{equation}
This result actually generalizes to any non-constant initial density as long as Eq.~\ref{expodep} holds, because of the interplay between the logarithm in the definition of the A$+$ indicator and the exponential dependence of Eq.~\ref{expodep}, which leads to the first and the third term in Eq.~\ref{aless_const} canceling out. Thus the A$+$ indicator should evolve linearly with time if the dynamical friction timescale is constant with radius. 

\section{Conclusions}
Working within a pure dynamical friction picture, under a set of simplifying assumptions, I have shown that the \citet{2016ApJ...833..252A} A$+$ indicator evolves monotonically in time and I have found an analytical solution for its time dependence. I worked out the case of a dynamical friction timescale that is constant with radius, which results in the A$+$ indicator increasing linearly with time. Monotonicity is an interesting result, as it proves that the A$+$ indicator is effectively a \emph{dynamical clock} as previously claimed by \citet{2016ApJ...833..252A} based on the results of a set of direct N-body simulations. As my simple model neglects diffusion, which was instead treated numerically by \citet{2018ApJ...867..163P}, I showed that the A$+$ indicator still works as a dynamical clock even in the absence of diffusion.

\section*{Acknowledgments}
This project has received funding from the European Union's Horizon $2020$
research and innovation programme under the Marie Sk\l{}odowska-Curie grant agreement No. $664931$.
I wish to thank Dr. Pierfrancesco di Cintio, Dr. Paolo Miocchi, Dr. Alessandro Cobbe, and Dr. Stefano Pugnetti for helpful discussion on this subject.

\bibliographystyle{rmaa}
\bibliography{ms}

\end{document}